\documentclass[prl, twocolumn, superscriptaddress,showpacs]{revtex4}

\usepackage{bm}
\usepackage{epsfig}
\input epsf

\begin{document}

\title{ Quantum Hall-like effect on strips due to geometry}

\author {R. Dandoloff}
\email{rossen.dandoloff@ptm.u-cergy.fr}
\author{  T.T. Truong}

\affiliation{ Laboratoire de Physique Th\'{e}orique et
Mod\'{e}lisation, Universit\'{e} de Cergy-Pontoise, F-95302
Cergy-Pontoise, France}

\begin{abstract}
In this Letter we present an exact calculation of the effective
potential which appears on a helicoidal strip. This potential
leads to the appearance of localized states at a distance $\xi_0$
from the central axis. The twist $\omega$ of the strip plays the
role of a magnetic field and is responsible for the appearance of
these localized states and an effective transverse electric field
thus this is reminiscent of the quantum Hall effect. At very low
temperatures the twisted configuration of the strip may be
stabilized by the electronic states.
\end{abstract}


\maketitle
The appearance of nanostructures has boosted the
interest in quantum strip waveguides and in tubular quantum
waveguides. Of special interest is the appearance of bound states
in these structures. For a thin tubular waveguide the binding
potential is \cite{Goldstone} $
V_{eff}\approx-\frac{\hbar^2}{2m}\frac{k^2}{4}$ where $k$ is the
curvature of the axis of the tube, viewed as a space curve. In the
case of a quantum strip the effective potential is:\cite{Clark}
$$ V_{eff}\approx\frac{\hbar^2}{2m}\left(-\frac{k^2}{4}+\frac{1}{2}\left[\tau -
\theta_s\right]^2\right) \eqno(1) $$ where $\tau $ is the torsion
of the strip axis and $\theta$ is the twist angle around the axis.
Here the subscript $s$ stands for $\frac {d}{ds}$ where $s$ is the
arc-length of the curve. Usually these results are valid for thin
tubes and narrow strips and the curvature $k$ should be
small.\cite{Goldstone}. The curvature $k$ is {\it responsible} for
the appearance of bound states in both types of waveguides.

In this Letter we evaluate the {\it sole} effect of twisting of a
strip and show that a pure twist may cause localization and play
the role of an applied magnetic field.

We consider a strip whose edge is a straight line along the
$x$-axis and whose other edge follows a helix around the $x$-axis.
The surface represents a helicoid and is given by the following
equation:
$$
{\mathbf r}=x\, {\mathbf e}_{x}+ \xi\, [\cos(\omega x)\, {\mathbf
e}_{y}+ \sin(\omega x)\, {\mathbf e}_{z}], \eqno(2)
$$
where $\omega = \frac{2\pi n}{L}$, $L$ is the total length of the
strip and $n$ is the number of $2\pi$-twists.

\begin{figure}
\begin{center}
   \epsfysize= 5cm
   \epsfbox{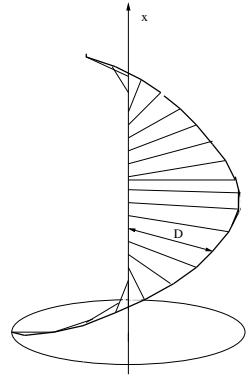}
   \caption{\large {Helicoidal Surface}}
\end{center}
\end{figure}

$(\mathbf{e}_x,\mathbf{e}_y,\mathbf{e}_z )$ is the usual
orthonormal triad in $R^3$ and $\xi\in [0, D]$, where $D$ is the
width of the strip. Let $\mathbf{ dr}$ be the displacement
$$  {\mathbf dr} = dx\, {\mathbf e}_x + [\cos(\omega x)\,{\mathbf e}_y
+ \sin (\omega x)\,{\mathbf e}_z]\,d\xi + $$
$$+[-\omega\xi\sin(\omega x)\,{\mathbf e}_y + \omega\xi\cos(\omega
x)\,{\mathbf e}_z]\,dx \eqno(3)
$$
and therefore we have:
$$
|{\mathbf dr}|^2 = (1 + \omega^2\xi^2)dx^2 + d\xi^2 = h_1^2dx^2 +
h_2^2d\xi^2, \eqno(4)
$$
where $h_1=\sqrt{1+\omega^2\xi^2}$ and $h_2=1$ are the Lam\'e
coefficients of the induced metric (from $R^3$) on the strip. Now
the Hamiltonian for a free particle on the strip is given by:
$$
H=-\frac{\hbar^2}{2m}\frac{1}{h_1}\left[
\left(\frac{\partial}{\partial \xi}h_1\frac{\partial}{\partial
\xi}\right)+ \frac{\partial}{\partial x
}\frac{1}{h_1}\frac{\partial}{\partial x}\right] \eqno(5)
$$
After rescaling the wave function $\psi \mapsto \frac{1}{\sqrt
h_1}\psi$ (because we require the wave function to be normalized
with respect to the area element $dxd\xi$) and after some algebra
we arrive at the following expression for the Hamiltonian:
$$
H=-\frac{\hbar^2}{2m} \left[ -\frac{1}{2h_1} \left(
\frac{\partial^2 h_1}{\partial \xi^2}\right)+
\frac{1}{4}\frac{1}{h_1^2}\left( \frac{\partial h_1}{\partial
\xi}\right)^2+ \frac{\partial^2}{\partial \xi^2} \right]
$$
$$
-\frac{\hbar^2}{2m} \frac{1}{h_1^2}\frac{\partial^2}{\partial
x^2}. \eqno(6)
$$
This Hamiltonian may be rewritten in a more transparent form:
$$
H=-\frac{\hbar^2}{2m}\frac{\partial^2}{\partial \xi^2} +
V_{eff}(\xi)
-\frac{\hbar^2}{2m}\frac{1}{h_1^2}\frac{\partial^2}{\partial x^2},
\eqno(7)
$$
where the effective potential in the $\xi$ direction is given by:
$$ V_{eff}(\xi)=-\frac{\hbar^2}{2m} \left[
-\frac{1}{2h_1}\left( \frac{\partial^2 h_1}{\partial \xi^2}\right)
+ \frac{1}{4}\frac{1}{h_1^2}\left(\frac{\partial h_1}{\partial
\xi}\right)^2\right] $$

Note that in \cite{Goldstone} and \cite{Clark} the effective
potential is longitudinal. In the present case there is no
longitudinal effective potential. After insertion of
$h_1=\sqrt{1+\omega^2\xi^2}$ the effective potential becomes:
$$
V_{eff}(\xi)=\frac{\omega^2\hbar^2}{4m}\frac{1}{(1+\omega^2\xi^2)^2}\left[1-\frac{\omega^2\xi^2}{2}\right].
\eqno(8)
$$
This effective potential is of pure quantum-mechanical origin
because it is proportional to $\hbar$. Note that this expression
is exact and is valid not only for small $\xi$: here no expansions
in small parameter has been used. On the axis of the helicoid
$\xi=0$ we get the following value of the repulsive potential $
V_{eff}(0)=\frac{\omega^2\hbar^2}{4m}$ which corresponds to the
expression for the effective potential found in \cite{Clark} for
$k=0$ and $\tau=0$ and $\theta_s=\omega$. $V_{eff}(0)$ represents
a local maximum of the potential. The local minimum is reached for
$\xi_0=\frac{\sqrt 5}{2\pi}\frac{L}{n}$ and
$V_{eff}(\xi_0)=-\frac{1}{24}\frac{\omega^2\hbar^2}{4m}.$

Now we may write the time-independent Schr\"odinger equation as:
$$
\left[-\frac{\hbar^2}{2m}\frac{\partial^2}{\partial \xi^2} +
V_{eff}(\xi)\right]\psi -
\frac{\hbar^2}{2m}\frac{1}{h_1^2}\frac{\partial \psi}{\partial
x^2}=E\psi \eqno(9)
$$
Using the ansatz: $\psi(x,\xi)=\phi (x)f(\xi)$:
$$
\left[-\frac{\hbar^2}{2m}\frac{h_1^2(\xi)}{f(\xi)}\frac{\partial^2f(\xi)}{\partial
\xi^2}+(V_{eff}(\xi)-E)h_1^2(\xi)\right]
$$
$$
-\frac{\hbar^2}{2m}\frac{1}{\phi (x)}\frac{\partial^2 \phi
(x)}{\partial x^2}=0 \eqno(10)
$$
we get two differential equations:
$$
\frac{\hbar^2}{2m}\frac{d^2 \phi(x)}{dx^2}=-E_0\phi(x), \eqno(11)
$$ and
$$
-\frac{\hbar^2}{2m}\frac{d^2f(\xi)}{d\xi^2}+\left[V_{eff}(\xi)+\frac{E_0}{h_1^2(\xi)}\right]f(\xi)=E.
\eqno(12)
$$
With $\phi(x)=e^{ik_xx}$ in eq(11) we have
$E_0=\frac{k_x^2\hbar^2}{2m}$ where $k_x$ is the partial momentum
in $x$-direction.

It is clear from eq(8) that for $\xi$ close to 0, $V_{eff}$
represents a repulsive potential and the twist $\omega$ works
"against" the appearance of localized states \cite{Clark}.
However, for $\xi \ge \frac{\sqrt 2}{\omega}=\frac{\sqrt
2}{2\pi}\frac{L}{n}=\xi_0$ (there are no restrictions on $\xi$ in
eq.(8)) , $V_{eff}\le 0$ and there will be localized states.
Physically one may understand the appearance of localized states
away from the central axis using the Heisenberg uncertainty
principle: for greater $\xi$ a particle on the strip will dispose
with more "space" along the corresponding helix and therefore the
corresponding momentum and hence energy will be smaller than for a
particle closer to the central axis. Thus the twist $\omega$ will
"push" the electrons towards the outer edge of the strip and
create an effective electric field between the central axis and
the helix. The depth of the potential minimum depends on $\omega$.
Thus the number of localized states (and their existence) will
depend on the width of the strip and on the twist. The minimum of
the potential in eq(12) is reached for $\xi=\xi_0$ and is given
by:
$$ U_{min}=\frac {\hbar^2}{6m}\left[ k_x^2 -
\frac{\omega^2}{16}\right].  $$ For small $k_x$ i.e.
$k_x\le\frac{\omega}{4}$, $E_{min}\le 0$ and there is a
possibility for having localized states with negative energy
levels. For very low temperatures most of the $k_x's$ will be well
below $\frac{\omega}{4}$ ( $\frac{\hbar^2\omega^2}{2m}\sim k_BT $,
where $k_B$ is the Boltzmann constant and $T$ is the temperature).
For a strip of width at least $D=\xi_0$, the twisting the strip
will increase its elastic energy ( the elastic energy density per
unit length is $\frac{1}{2}C^{\ast}\omega^2$, where $C^{\ast}$ is
the torsional constant) but on the other hand will create
localized states with negative energy levels which will diminish
the total electronic energy and for rather soft materials it may
favor the twisted configuration against that of the flat strip.

Eq(12) represents the motion in the direction $\xi$ with a net
potential
$$ U(\xi)=\frac{\omega^2}{4}\left\{
\frac{4C^2-1}{(1+\omega^2\xi^2)} + \frac{3}{(1+\omega^2\xi^2)^2}
\right\} \eqno(13) $$ where $C=\frac{k_x}{\omega}$. This potential
is a sum of two contributions, a repulsive part:
$\frac{3}{(1+\omega^2\xi^2)^2}$ and a variable part which is
repulsive for $C^2\geq \frac{1}{4}$ and attractive for $C^2\le
\frac{1}{4}$. If $C\le\frac{1}{2}$ the $U(\xi)$ becomes negative
for $\omega\xi\geq\sqrt{\frac{2+4C^2}{1-4C^2}}$ and one expects
bound states with negative energy eigenvalues in this potential
well. The finite size of the width $D$ determines the cut-off of
$U(\xi)$ and hence the probability for a particle to be "pushed"
to the boundary of the strip.

Equation of motion (12) may take the remarkable normal form of the
equation of a confluent Heun function\cite{Heun}: let us call
$f(\xi)=H(\omega^2\xi^2)=H(z)$. Then putting
$e=\frac{\epsilon}{4\omega^2}$, we get an equation for the
function $H(z)$:
$$
-zH''(z)-\frac{1}{2}H'(z)+\frac{1}{16}\left\{ \frac{4C^2-1}{1+z}+
\frac{3}{(1+z)^2}\right\}H(z)=
$$
$$
-eH(z). \eqno(14)
$$
A further change of function $H(z)=z^{1/4}L(z)$ leads to:
$$ -zL''-\frac{3}{16}L+\frac{1}{16}\left\{ \frac{4C^2-1}{1+z} +
\frac{3}{(1+z)^2}\right\}L=-eL. \eqno(15) $$ Now define
$\zeta=1+z$ and set $L(z)=M(\zeta)$, the equation satisfied by
$M(\zeta)$ is of the form:
$$ M''(\zeta)+Q(\zeta)M(\zeta)=0. \eqno(16) $$
with
$$ Q(\zeta)=-\left(
e+\frac{4C^2-1}{16}\right)\frac{1}{\zeta-1}+\frac{4C^2+2}{16\zeta}+\frac{3}{16\zeta^2}
\eqno(17) $$ This is to be compared to the normal form of a
confluent Heun equation:
$$ y'(x)+\left\{
A+\frac{B}{x}+\frac{C}{x-1}+\frac{D}{x^2}+\frac{E}{(x-1)^2}\right\}y(x)=0.
$$
Thus eq(16) is really a confluent Heun equation with
\[A=0,\,\,\,
B=\frac{4C^2+2}{12}, \,\,\,
C=-\left(e+\frac{4C^2-1}{16}\right),
\]
\[\,\,\, D=\frac{3}{16},\,\,\,
E=0
\]
which all depend on $C=\frac{k_x}{\omega}$ representing the ratio
of straight propagation in the $x$-direction over the geometric
twist. As the properties of confluent Heun functions are not
extensively known, e.g. zeros have not yet been completed, we
shall not dwell on it.

\end{document}